\documentclass[conference]{IEEEtran} 

\usepackage{cite}
\usepackage[pdftex]{graphicx}
\usepackage{url}
\usepackage{blindtext}
\usepackage{hyperref}
\usepackage{balance}
\usepackage{soul,color}
\soulregister\cite7
\soulregister\ref7
\soulregister\pageref7
\usepackage{enumitem}
\usepackage{pifont}
\usepackage{url}
\usepackage{amsmath,amssymb,amsfonts}
\usepackage{algorithmic}
\usepackage{textcomp}
\usepackage{xcolor}
\usepackage{booktabs}
\usepackage{epigraph}
\usepackage{pifont}

\begin{document}
\title{From Failure to Alignment: A Requirements Engineering Framework for Machine Learning Systems}

\author{
\IEEEauthorblockN{
Amel Bennaceur\IEEEauthorrefmark{1},
Gopi Krishnan Rajbahadur\IEEEauthorrefmark{2},
Prince Mercy\IEEEauthorrefmark{3},
Bashar Nuseibeh\IEEEauthorrefmark{1}\IEEEauthorrefmark{4},
and Faeq Alrimawi\IEEEauthorrefmark{3}
}

\IEEEauthorblockA{\IEEEauthorrefmark{1}The Open University, United Kingdom}

\IEEEauthorblockA{\IEEEauthorrefmark{2}Queen's University, Canada}

\IEEEauthorblockA{\IEEEauthorrefmark{3}University of Limerick, Ireland}
\IEEEauthorblockA{\IEEEauthorrefmark{1}City St George's, United Kingdom}

}

\maketitle

\begin{abstract}
Organisations designing, developing, and deploying machine learning systems (MLS) need to be able to check that these systems are trustworthy, and communicate this clearly to their stakeholders, be they different categories of users, engineers, or wider society. By focusing on stakeholders, Requirements Engineering is well positioned to drive the design and engineering of MLS that align with the needs of their stakeholders. Yet, we still need a systematic process for modelling and reasoning about requirements for MLS that is driven both by stakeholders' needs and constraints for MLS development.

This paper proposes a framework entitled REAL (Requirements Engineering for mAchines that Learn - and Fail) to help develop MLS that align with stakeholders' needs by adopting a requirements engineering approach. This model-based framework is based on three principles. First, weaving together requirements for data, models, and the system as a whole. Second, using failure to drive the exploration of alternative requirements. Third, iterative and traceable refinement of MLS requirements.

We demonstrate the proposed framework using an example from autonomous driving and show that REAL supports the development of MLS that better align with stakeholders' requirements. A replication package is available \href{https://github.com/darkaengl/REAL}{online}.
\end{abstract}

\IEEEpeerreviewmaketitle

\section{Introduction}
\label{sec:intro}
\epigraph{“I have not failed. I've just found 10,000 ways that won't work.”}{Thomas A. Edison}

Modern software-intensive systems increasingly incorporate
Machine Learning (ML) components to enable data-driven
decision-making. Engineering such Machine Learning Systems (MLS),
however, differs from traditional software
engineering in several ways: behaviour is learned rather than explicitly
programmed, performance is distribution-dependent, and failure
is inherent due to generalisation and uncertainty~\cite{kastnerTeachingSoftwareEngineering2020, amershiSoftwareEngineeringMachine2019}.
Underspecification further challenges the reliability of deployed
models~\cite{DAM20}. While industrial best practices are emerging~\cite{bestPracticesGovAI23},
systematic techniques to specify, analyse, and assure requirements
for MLS remain limited~\cite{AhmadAABG23}.

Requirements Engineering (RE) provides mechanisms to capture
stakeholder needs and reason about the relationship between
domain assumptions, system specifications, and requirements~\cite{BennaceurTYN19, ZaveJ97}.
For MLS, this relationship becomes conditional and evolving:
the operational domain is open-ended, environmental variability
is vast, and complete
upfront specification is unrealistic.
The operationalisation of requirements
introduces assumptions about sensing limits, data coverage,
and model behaviour and requirement satisfaction must 
be evaluated across domain instantiations.
This raises the following questions:

\begin{itemize}[leftmargin=*]
    \item {RQ1:} How can reasoning about requirements  be operationalised for MLS?
    \item {RQ2:} How can failure discovery and mitigation be systematically integrated into the RE process?
    \item {RQ3:} To what extent does failure-driven, cross-layer adaptation measurably improve requirement satisfaction?
\end{itemize}

While many industrial guidelines emphasise the need to deal with failure in MLS, this has received little attention in RE research~\cite{failureModes19}. Failures are an integral part of MLS development because of generalisation~\cite{failureModes19}.
Detecting failures has received significant attention in ML Testing~\cite{ZhangHML22}.
However, not all failures are the same and we need to understand which ones stakeholders can tolerate and which ones are unacceptable~\cite{JodatCNS24}.
In particular, understanding the operational design domain is paramount to specify and scope MLS, especially in safety-critical contexts \cite{FahmyPBS23,mattioli2023ai}. 

We propose to treat failure as a first-class
construct and use it to expose hidden assumptions,
surface trade-offs, and guide
adaptation across data, model, system, and requirement layers.
Counterexamples become diagnostic artefacts linking empirical
behaviour to stakeholder requirements.
To operationalise this perspective, we introduce \textit{REAL
(Requirements Engineering for mAchines that Learn - and Fail)}, a
failure-driven RE framework for MLS.
REAL embeds grammar-guided scenario exploration, obstacle
analysis, and cross-layer mitigation within an iterative
alignment loop that reconnects testing outcomes to requirement
and domain refinement.

There has been extensive work on RE for Artificial Intelligence (AI) and MLS in recent years~\cite{YoshiokaHTCWF21, PeiLWW22, AhmadAABG23, VillamizarEK21}. For example, R4ML~\cite{nalchigarModelingMachineLearning2021} is a conceptual modelling framework for requirements elicitation,
design, and development of ML solutions that relates business requirements view with analytics design and data preparations views.
Chuprina \textit{et al.}~\cite{chuprinaArtefactbasedRequirementsEngineering2021} propose a multi-model approach whereby the requirements and data models are explicitly linked through dependency relationships.
However, it is unclear how those links evolve or are adapted as requirements change or failures are discovered. 
The conceptual model proposed by Ahmad \textit{et al.}~\cite{AhmadAABBG23} is the closest to ours. It proposes to help engineers elicit and specify requirements for MLS by modelling multiple needs, namely users, models, data, feedback \& user control, explainability \& trust, errors \& failures. 
REAL also proposes to connect those concepts through the analysis of failure and its relationship to obstacle analysis. Furthermore, we propose to also use mitigation techniques within the feedback loop.
Anunnaki~\cite{AnuCheng24} also relies on goal modelling and obstacle analysis to adapt MLS to the uncertainty of the environment. However, it does not use failure to drive the adaptation of the requirements themselves.
Finally, unlike approaches that treat failures as artefacts of testing or model debugging\cite{
ZhangHML22,FahmyPBS23,JodatCNS24}, REAL formalises failure as a counterexample to the satisfaction relation and integrates obstacle analysis and cross-layer mitigation within a unified requirements refinement loop.
This paper makes the following contributions:

\begin{itemize}[leftmargin=*]
    \item \textit{Operational linkage between scenario generation and requirements analysis.}
    We connect stakeholder requirements to executable domain
    variability through grammar-guided scenario exploration,
    enabling systematic discovery of requirement-relevant
    counterexamples within admissible operational bounds.

    \item \textit{Obstacle-based interpretation of MLS failures.}
    We reinterpret empirical failures as obstacles in a KAOS-style
    goal model, providing a principled mechanism to diagnose
    misalignment between stakeholder goals, domain assumptions,
    and system behaviour.

    \item \textit{Multi-layer mitigation strategy.}
    We define and evaluate coordinated adaptation across data,
    model, system, and requirement layers, making explicit
    how mitigation decisions affect requirement satisfaction.

    \item \textit{Autonomous driving demonstrator.}
    We demonstrate REAL on an autonomous braking system
    using CARLA, Scenic, and evolutionary scenario generation,
    and provide a \href{https://github.com/darkaengl/REAL}{publicly available} replication package to
    support reproducibility.
\end{itemize}

The remainder of the paper is structured as follows.
Section~\ref{sec:example} introduces the autonomous braking example.
Section~\ref{sec:background} presents background on goal modelling.
Section~\ref{sec:framework} details the REAL framework.
Section~\ref{sec:validation} reports the results of our evaluation using the autonomous braking example.
Section~\ref{sec:relatedWork} discusses related work.
Finally, Section~\ref{sec:conclusion} concludes the paper and outlines future directions.

\section{Example: Automated Braking System}
\label{sec:example}
\begin{figure}[htbp]
\centering
\includegraphics[width=1.0\linewidth]{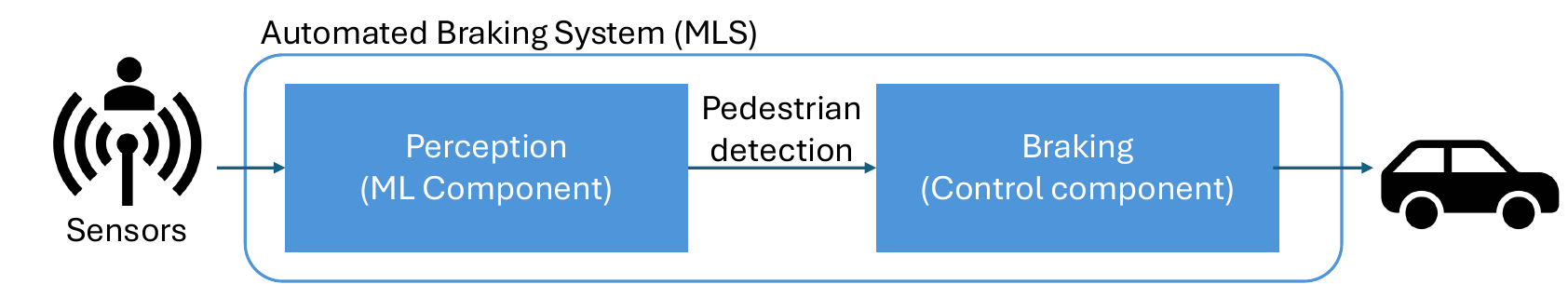}
\caption{An automated braking system example}
\label{fig:example}
\end{figure}

To ground the discussion and empirically evaluate REAL, we consider an automated braking system as an example of safety-critical MLS. Autonomous driving systems integrate sensing, perception, planning, and control, where violations of requirements may have severe real-world consequences. Open-source platforms such as Apollo~\cite{apollo} and Autoware~\cite{autoware} adopt modular architectures in which loosely coupled components communicate via well-defined message interfaces.  For clarity and focus, we isolate a minimal configuration comprising two components (Fig.~\ref{fig:example}): 
a \emph{pedestrian detection} component implemented using a machine learning model, and 
a \emph{braking controller} implemented using deterministic control logic.

The perception component receives sensor inputs (e.g.~image data), processes them using a trained object detection model, and outputs pedestrian classifications with associated confidence levels. The braking component activates the braking controller when a pedestrian is detected within a predefined safety threshold. While the braking logic is deterministic, the perception component is data driven and subject to distributional variation and environmental uncertainty.

From an RE perspective, this system allows us to reason explicitly about the relationship between:
\begin{itemize}[leftmargin=*]
    \item stakeholder safety requirements (e.g.~the vehicle shall avoid collisions with pedestrians),
    \item assumptions about the driving environment (e.g.~visibility conditions and pedestrian characteristics), and
    \item the implemented perception and control components.
\end{itemize}

In practice, variations in pedestrian size, clothing, lighting, or weather conditions may lead to situations in which the system fails to satisfy safety requirements. Such failures  represent misalignments between stakeholder expectations, domain assumptions, and system behaviour.
The example provides a controlled yet realistic setting to analyse requirement satisfaction of MLS. First, it exposes the interaction between learned and non-learned components. Second, safety requirements are explicit and operationalisable within a simulator. Third, the operational design domain can be parameterised along dimensions such as pedestrian attributes, distance, direction, and environmental conditions, allowing systematic exploration of diverse scenarios. 
The following section introduces the foundational concepts in goal-oriented RE.

\section{Background: Goal-oriented RE and KAOS}
\label{sec:background}

Zave and Jackson~\cite{ZaveJ97} define three main artefacts produced during the RE process: 
\emph{domain properties}, which describe assumptions about the environment that hold independently of the machine; 
\emph{requirements}, which prescribe properties that stakeholders wish to hold in the presence of the machine; and 
\emph{specifications}, which prescribe what the machine must do to ensure that those requirements are achieved.

\begin{center}
Under domain assumptions ($D$), the specification ($S$) 
should satisfy the requirements ($R$): $D, S \models R$.
\end{center}

This separation between domain assumptions and machine specification is critical: if domain properties are incomplete, unrealistic, or violated in practice, requirement satisfaction may fail even when the specification appears correct.
This distinction becomes particularly important for MLS where behaviour is learned from data and therefore depends implicitly on assumptions about training distributions, operational environments, and feature representations. Failures often arise not only because the implementation is flawed, but because domain assumptions are underspecified, violated, or evolve over time. Making these assumptions explicit is therefore a prerequisite for systematic reasoning about requirement satisfaction under uncertainty.

Multiple modelling techniques can be used. REAL builds on
goal modelling techniques that focus on \textit{why}, i.e., the rationale and objectives of the different system components or actors as well as \textit{who} is responsible for realising them.
In particular, REAL uses KAOS~\cite{Lamsweerde08} which focuses on refinement relationships.

\begin{figure}[t]
\centering
\includegraphics[width=1.0\linewidth]{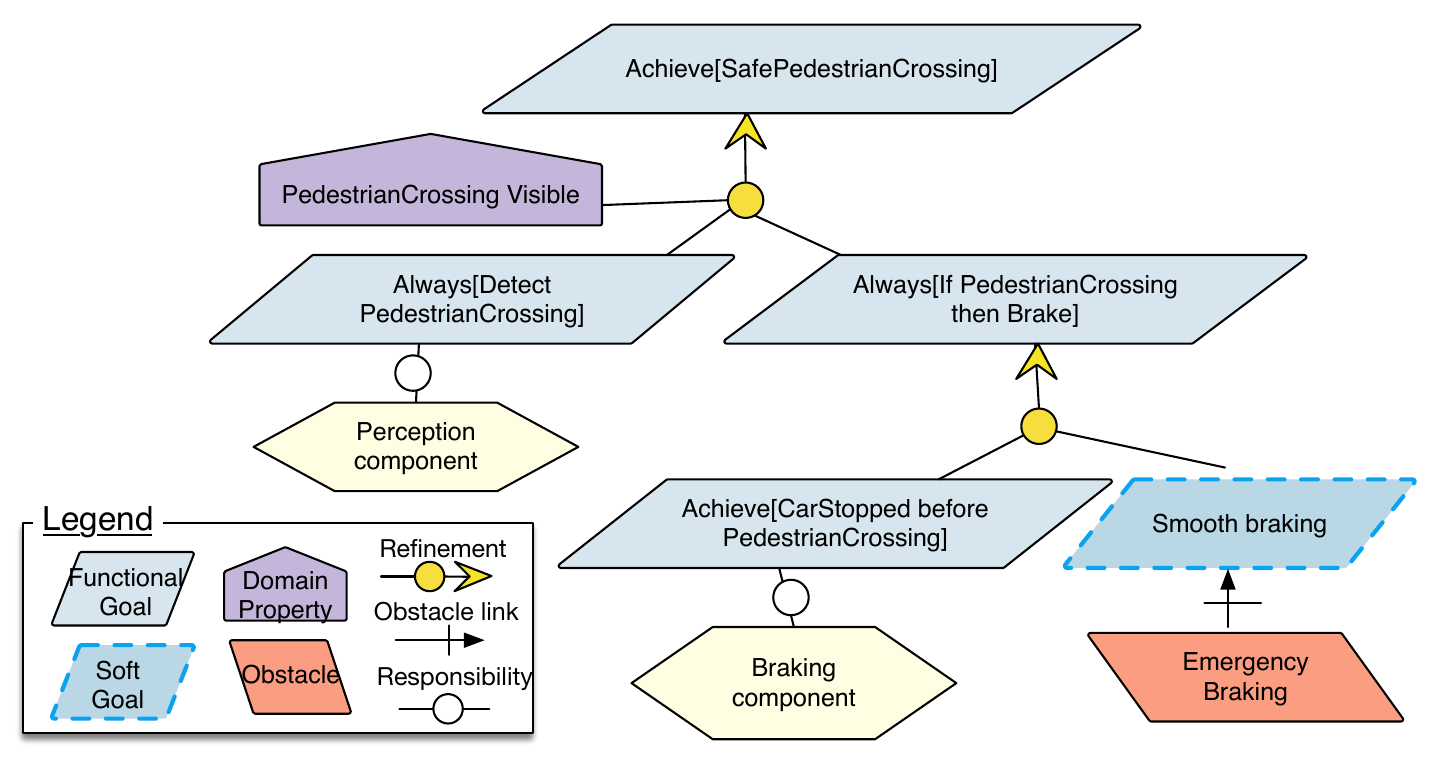}
\caption{A sample KAOS model for safe pedestrian crossing}
\label{fig:kaos}
\end{figure}

Let us consider an example of \emph{safe (autonomous) braking at a pedestrian crossing} illustrated in Fig.~\ref{fig:kaos}. 
The high-level goal \emph{Achieve[SafePedestrianCrossing]} captures a stakeholder requirement: pedestrians should be able to cross safely. This goal is refined into sub-goals and supported by domain properties.
A \emph{goal} is a prescriptive statement describing a desired property of the system. Goals may be
(i)~\emph{functional}, describing required services (e.g.~\emph{Always[DetectPedestrianCrossing]}), or
(ii)~\emph{soft}, describing quality attributes subject to optimisation (e.g.~\emph{SmoothBraking}).
Domain properties describe assumptions about the environment. For example, \emph{PedestrianCrossingVisible} expresses the assumption that the crossing is perceptible by the system. The satisfaction of the high-level safety goal depends jointly on sub-goals and domain properties: safe crossing can be achieved if the system consistently detects the crossing and brakes appropriately, provided the crossing is visible.
Refinement links decompose high-level goals into more operational sub-goals and enable the allocation of responsibility to agents. For instance, the \emph{Perception component} is responsible for satisfying \emph{Always[DetectPedestrianCrossing]}, while the braking component is responsible for operationalising \emph{Always[If PedestrianCrossing then Brake]}. Through refinement, stakeholder intentions are progressively translated into implementable and verifiable behaviours.
KAOS also supports reasoning about risks through the identification of obstacles, i.e.~conditions that may prevent goals from being satisfied. An obstacle captures a situation in which the current specification and domain assumptions are insufficient to ensure goal satisfaction. For example, \emph{EmergencyBraking} may prevent the satisfaction of the soft goal \emph{SmoothBraking}.

\begin{figure*}[t]
\centerline{\includegraphics[width=1.0\linewidth]{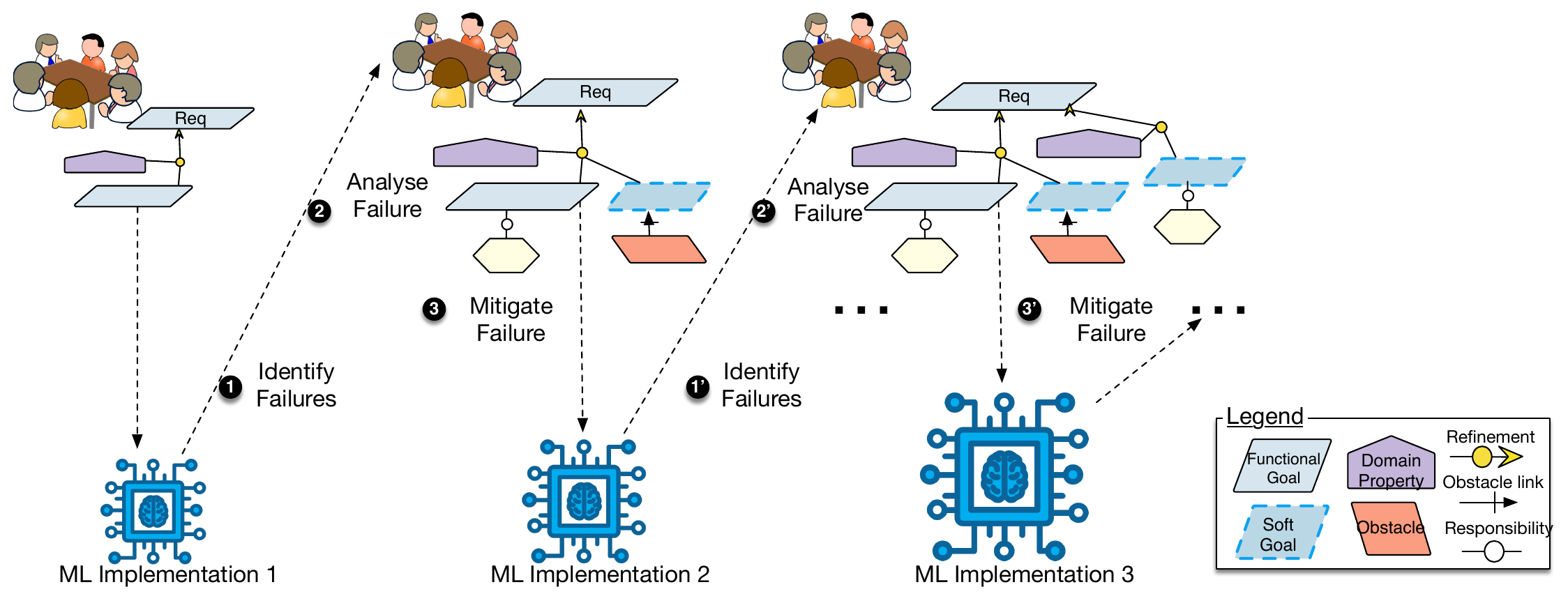}}
\caption{Overview of the REAL Framework}
\label{fig:process}
\end{figure*}

\section{The REAL Framework}
\label{sec:framework}
The relation $D, S \models R$ captures alignment between stakeholder requirements $R$ and system behaviour $S$ under stated environmental assumptions $D$. For MLS, $S$ is often underspecified as behaviour is learned from data rather than specified, and domain assumptions are often implicit, distribution-dependent, or only partially understood. A system may satisfy requirements under nominal assumptions $D$ yet violate them under alternative admissible domain instantiations $D'$ ($D', S \not\models R$).
These violations correspond to observable failures. REAL adopts the position that failure is not exceptional but expected in MLS, and should be treated as diagnostic evidence of misalignment between domain assumptions, requirements, and implementation.
REAL begins with stakeholder elicitation and formulation of high-level goals $R$. These goals are refined into an initial specification $S$, encompassing data selection, model configuration, and system integration. This early implementation operationalises the current understanding of domain assumptions $D$.

Failure identification (Fig.~\ref{fig:process}-\ding{182}) is performed through structured scenario exploration, simulation, and testing. REAL searches for admissible domain instantiations under which the current specification fails to satisfy stakeholder requirements. Each discovered failure provides evidence that: (i) domain assumptions are incomplete; (ii) requirements are underspecified or unrealistic; or (iii) the specification does not adequately operationalise the requirements.

Failure analysis (Fig.~\ref{fig:process}-\ding{183}) considers the discovered failures as structured evidence of misalignment between stakeholder goals, domain assumptions, and system specification. Failures are organised and grouped using trace classification or clustering to reveal recurring behavioural patterns, enabling systematic obstacle identification rather than ad hoc debugging. 
Through this process, empirical counterexamples are elevated into explicit obstacle conditions that guide subsequent refinement of requirements, domain assumptions, or implementation decisions.

Failure mitigation (Fig.~\ref{fig:process}-\ding{184}) operationalises repair within the REAL alignment loop by distinguishing three types of adaptation. 
\emph{Specification adaptation} revises the implementation to obtain $S'$, for example through changes to data, model configuration, decision thresholds, or system control, 
while keeping requirements unchanged, with the aim of restoring satisfaction under the same $R$, that is $D', S' \models R$. 
Second, \emph{requirement refinement} yields a revised requirement set $R'$ that more precisely captures stakeholder needs, i.e.~$D', S \models R'$. Third, \emph{domain revision} makes previously implicit environmental assumptions explicit, producing revised domain properties $D''$ clarifying the boundaries of the Operational Design Domain (ODD) under which satisfaction can be defensibly reasoned about $D'', S \models R$. 
These mitigation actions may occur across data, model, system, and requirement layers, but are always evaluated against an explicitly scoped domain. After adaptation, the satisfaction relation is re-assessed; if additional admissible domain instantiations expose new violations, the Identify–Analyse–Mitigate cycle iterates.

This process yields an iterative alignment loop in which violating domain instantiations are identified, misalignment between domain assumptions, requirements, and specification is analysed, the relevant artefacts ($S$, $R$, or $D$) are adapted accordingly, and satisfaction is subsequently re-evaluated.
Unlike approaches that assume fixed domain assumptions and static requirements, REAL treats $D$, $R$, and $S$ as evolving artefacts embedded within a structured, failure-driven refinement process.
In the following, we detail each phase of the framework.

\subsection{Identifying Failure through Scenario Exploration}

For MLS, behaviour is learned from data and is distribution-dependent. Consequently, requirement satisfaction cannot be assumed uniform across the operational domain but must instead be evaluated over concrete domain instantiations.
We operationalise the domain as a parametrised scenario space 
$D(\theta), \theta \in \Theta$
where parameters $\theta$ encode environmental variables (e.g.~pedestrian size, crossing direction, or weather conditions). This representation makes explicit the variability that is often only implicitly assumed during development.

To systematically explore $\Theta$, REAL adopts a grammar-based scenario generation using Grammatical Evolution (GE)~\cite{ge}. GE is an evolutionary algorithm that automatically generates programs using a Domain Specific Language (DSL) defined by a formal grammar. Initially introduced as an automatic programming tool, GE can produce candidate scenarios for domains described using a Backus–Naur Form (BNF) grammar. 
The grammar defines admissible combinations of actors, environmental conditions, and behaviours"

\[
\langle scenario \rangle ::= 
\langle actor \rangle \;
\langle environment \rangle \;
\langle behaviour \rangle
\]

 The  grammar makes explicit the domain factors already present in the goal model and experimental design. Iterations may refine this grammar by adding new actors, behaviours, or constraints surfaced through stakeholder analysis or observed failures.
This construction enables controlled exploration of the search space and the discovery of rare or boundary combinations of domain factors. Generated scenarios are executed in simulation, yielding traces:
$Sim(D(\theta)) \rightarrow \textit{trace}$.

A failure is observed when the resulting execution trace violates a requirement:
$\textit{trace} \not\models R$.
Not all violations, however, correspond to meaningful requirement failures. Some arise from unrealistic or out-of-scope domain configurations. REAL therefore distinguishes between \emph{valid failures}, which occur under admissible domain assumptions and reflect genuine misalignment between requirements and system behaviour, and \emph{spurious failures}, which are triggered by scenarios that fall outside the intended operational design domain and thus indicate breaches of environmental assumptions rather than requirement violations.
Let $\Phi_{\text{valid}}(D(\theta))$ denote admissibility constraints encoding physical plausibility, safety bounds, and explicitly scoped operational assumptions. A failure is propagated to requirements analysis only when both requirement violation and admissibility hold:
\[
D(\theta), S \not\models R
\quad \text{and} \quad
\Phi_{\text{valid}}(D(\theta)) = \text{true}
\]

REAL is not prescriptive regarding the specific search mechanism (manual, statistical, evolutionary, or hybrid), but requires systematic exploration of the different attributes of the domain and the distinction between valid and spurious failures.
GE provides a structured and traceable exploration of the scenario space: generated scenarios are syntactically valid by construction, can be mapped back to explicit domain attributes, and can be refined when new obstacle classes are identified. Crafting the initial grammar requires domain knowledge and stakeholder input and REAL treats the grammar as an evolving requirements artefact rather than a fixed test generator. 
Furthermore, failure identification in REAL is necessarily relative to the explored scenario space $\Theta$, the grammar used to generate scenarios, and the admissibility predicate $\Phi_{valid}$. We therefore do not claim completeness with respect to all possible failures in the real world. Rather, REAL systematically explores the domain properties that have been made explicit, and uses discovered failures to surface missing assumptions and additional scenario attributes. For example, the simplified experiments reported in this paper focus on pedestrian size, clothing, and weather in order to make the core concepts clear. However, the accompanying artefact supports richer scenario specifications, including additional actors such as pets, and behaviours such as running. In other words, REAL does not provide a one-shot guarantee that failure identification is complete, but a structured process for progressively refining requirements and assumptions as new domain knowledge becomes available.

\vspace{0.3em}\noindent
\textbf{Example.}
In the autonomous braking example, the high-level requirement
\emph{Always[DetectPedestrian]} relies on assumptions such as pedestrian visibility and operation within specified environmental bounds.
Using the grammar-based generator integrated with a driving simulator, we systematically explored structured variations over pedestrian size, crossing direction, clothing, and weather conditions. This exploration revealed concrete scenarios in which the vehicle failed to brake appropriately.
Observed failures included cases involving child pedestrians with small apparent size, left-to-right crossing trajectories, and degraded weather conditions. Failures were analysed to determine whether they occurred under admissible assumptions or reflected a breach of operational boundaries. For example, extreme fog indicates insufficient scoping of domain assumptions.
Grammar-based scenario generation enabled systematic coverage of structured domain variability while validity filtering ensured that only meaningful counterexamples inform requirements reasoning. 
Scenario exploration thus provides structured evidence for refining requirements and domain assumptions.

\subsection{Analysing Failure through Obstacle Analysis}

Existing responsible-AI toolchains~\cite{MicrosoftRAI,Anthropic,GoogleAIGuide} emphasise the detection and mitigation of model- or system-level failures, yet they typically treat errors as local artefacts of training or deployment. They do not systematically reconnect failures to the underlying requirements problem, that is, to the interplay between stakeholder goals $R$, domain assumptions $D$, and system specification $S$. REAL addresses this gap by interpreting failures as manifestations of obstacles to goal satisfaction in the sense of goal-oriented requirements engineering~\cite{LamsweerdeL00,LetierL25}.

A failure discovered during scenario exploration corresponds to a counterexample 
$D(\theta), S \not\models R$
for some admissible domain instantiation $\theta$. Obstacle analysis reinterprets such a counterexample as evidence of a condition $O$ such that:
$
D(\theta), S \models O
\quad \text{and} \quad
O \Rightarrow \neg R.
$

An obstacle therefore characterises a state of the world under which goal satisfaction is prevented.
%
%
Failures may be organised and grouped using clustering or trace-classification techniques to reveal recurring patterns, thereby supporting systematic obstacle identification rather than ad hoc debugging. As proposed by Alrajeh \textit{et al.}~\cite{AlrajehKLRU12}, obstacles may also be synthesised from positive (successful) and negative (failed) execution traces, enabling semi-automated derivation of candidate obstacle conditions.
Through this process, failure analysis becomes a structured requirements activity rather than a purely model-centric diagnostic exercise. Empirical counterexamples are elevated into explicit obstacle conditions that guide refinement of domain assumptions, requirements, or system specification. By embedding obstacle analysis within scenario-based failure exploration, REAL aligns with the distribution-sensitive and evolving nature of MLS and supports iterative refinement toward a more defensible satisfaction relation under a well-scoped operational domain.

\begin{figure}[t]
\centering
\includegraphics[width=1.0\linewidth]{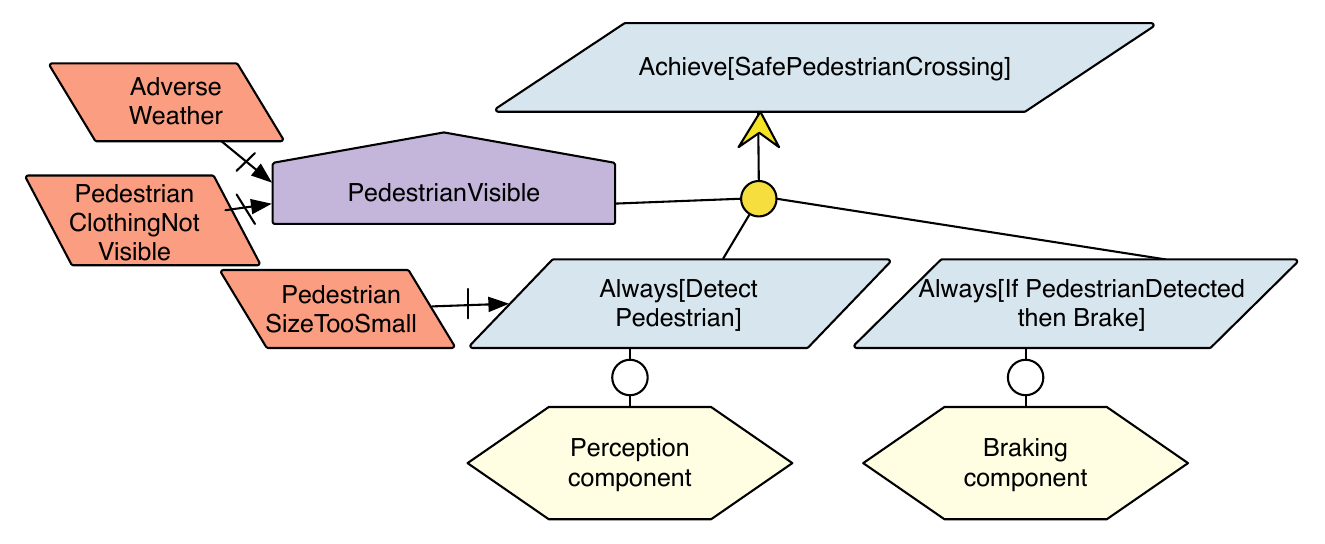}
\caption{Identifying obstacles for the automated braking example}
\label{fig:kaos_obs}
\end{figure}

\vspace{0.3em}\noindent
\textbf{Example.}
Consider the automated braking example illustrated in Fig.~\ref{fig:kaos_obs}. Scenario exploration identified failures involving child pedestrians under specific lighting and weather conditions.
The KAOS-style model (Fig.~\ref{fig:kaos_obs}) makes explicit that goal satisfaction relies on the domain property \emph{PedestrianVisible}. Failures involving adverse weather reveal that this assumption does not hold in certain domain instantiations. The obstacle \emph{PedestrianSizeTooSmall} further shows that  pedestrian characteristics, which might be related to sensor or model capabilities,  may also hinder the satisfaction of the \emph{Always[DetectPedestrian]} goal. A child pedestrian may be physically present and within the intended operational domain, yet have a small apparent size in the camera frame, reducing the likelihood that the perception component classifies the pedestrian with sufficient confidence. 
If the empirical distribution $P$ under-represents adverse visibility conditions or small-pedestrian instances, the learned predictor may violate the requirements under certain admissible domain instantiations. In this case, environmental conditions and pedestrian attributes are implicit parameters defining the operational design domain, and should be refined accordingly.
The failure therefore exposes a mismatch between: (i) the empirical distribution underlying training; (ii) the effective operational domain explored through scenario generation; and (iii) the safety requirement. Obstacle analysis separates the obstacle \emph{AdverseWeather}, which challenges the environmental visibility assumption, from \emph{PedestrianSizeTooSmall}, which challenges the detectability of a valid pedestrian instance under otherwise admissible conditions. Obstacle analysis elevates this mismatch from a local detection error to explicit obstacle conditions that can guide targeted mitigation, such as augmenting the training distribution with child-pedestrian scenarios, revising detectability assumptions, or adding confidence-aware braking behaviour.

\subsection{Mitigating Failure through Multi-Layer Adaptation}
Given a goal model specifying stakeholder requirements $R$, domain assumptions $D$, and an observed failure scenario $D(\theta)$ such that $D(\theta), S \not\models R$, the objective of this phase is to restore satisfaction through adaptation. 
In MLS, the specification $S$ includes training data, learned models, decision thresholds, and control logic. Consequently, mitigation must be understood as a \emph{multi-layer adaptation problem}. REAL distinguishes four layers of adaptation: data, model, system, and requirements.

\textit{Data-level adaptation.}
Failures may indicate that the empirical training distribution does not adequately approximate the operational domain in which the system is deployed. In such cases, mitigation must occur at the data level. This may involve augmenting the dataset with underrepresented or boundary scenarios, collecting additional samples from failure-inducing conditions identified during scenario exploration, rebalancing class distributions to reduce bias, or refining feature representations to better capture salient environmental attributes. Systematic error analysis techniques, such as slice-based analysis in the Microsoft Responsible AI Toolbox~\cite{MicrosoftRAI}, support identification of sub-populations associated with elevated error rates, while robustness literature emphasises targeted data augmentation to improve generalisation under distribution shift~\cite{GoodfellowSS14}. Formally, these interventions correspond to revising the empirical distribution $P$ from which training data $\mathcal{D}$ are drawn, yielding a modified distribution $P'$ such that the learned predictor $f' \in \mathcal{F}$ better approximates desired behaviour under operational domain instantiations $D(\theta)$. In this way, data-level adaptation seeks to restore the satisfaction relation by reducing the discrepancy between the training distribution and the effective deployment distribution.

\textit{Model-level adaptation.}
Failures may also stem from model configuration choices rather than insufficient data coverage. In such cases, mitigation occurs at the level of the learned model itself. Adaptation may involve adjusting decision thresholds to rebalance false positive and false negative rates, modifying architectural choices or hyperparameters, retraining with alternative optimisation objectives, or applying fine-tuning strategies such as transfer learning or few-shot updates. Robustness-oriented training methods, including adversarial training~\cite{GoodfellowSS14}, and systematic error analysis workflows such as those supported by the Microsoft Responsible AI Toolbox~\cite{MicrosoftRAI}, exemplify model-level mitigation practices that aim to reshape the learned decision boundary. Formally, these interventions transform the predictor $f \in \mathcal{F}$ into a revised mapping $f' : X \mapsto Y$, thereby altering the operational behaviour of the ML component while keeping stakeholder requirements $R$ and domain assumptions $D$ unchanged. In this sense, model-level adaptation seeks to restore satisfaction of the relation $D(\theta), S' \models R$ through modifications internal to the learned function rather than through revision of the requirements problem itself.

\textit{System-level adaptation.}
When the ML component exhibits limitations, system-level strategies can compensate for these limitations without modifying the learned model itself. Rather than retraining the predictor, adaptation can be achieved by modifying how it is embedded within the broader system architecture. Such mitigations include  (i) introducing confidence-aware control logic, where actions depend on prediction uncertainty; (ii) adding fail-safe behaviours that default to conservative actions under ambiguity; (iii) incorporating redundancy; and (iv) adjusting actuation policies to account for residual risk. These strategies are consistent with safety engineering practices in cyber-physical systems, which advocate runtime monitoring and adaptive assurance~\cite{AshmoreCP21}. At this level, adaptation results in a revised specification $S'$ such that requirement satisfaction may be restored under the same domain instantiation, i.e.,~$D(\theta), S' \models R$, demonstrating that alignment can be achieved through architectural integration and control redesign even when the ML component itself remains unchanged.

\textit{Requirement-level adaptation.}
Some failures do not necessarily indicate deficiencies in the implementation but instead expose underspecified requirements $R$ or incomplete domain assumptions $D$. In such cases, mitigation requires revisiting the satisfaction relation itself. When a failure scenario $D(\theta)$ yields $D(\theta), S \not\models R$ because assumptions embedded in $D$ are overly coarse or because $R$ lacks operational precision, refinement must occur at the level of $R$ or $D$ rather than solely at the level of $S$. This refinement process is fundamentally human-driven: stakeholders and engineers interpret the semantic meaning of failures, assess their acceptability, and decide whether assumptions, ODD boundaries, or goal formulations must be revised. However, it can be systematically supported by automated failure analysis techniques. In particular, clustering and trace classification methods such as the validity-aware labelling and graph-based failure mode clustering proposed in \textsc{Dynasto}~\cite{dynasto}, can group similar failure traces into interpretable modes, distinguishing systematic weaknesses from isolated outliers and separating valid failures from unrealistic scenarios. Such structured classification helps identify recurring patterns (e.g.~undetected children, consistent errors when a pedestrian is traversing left to right, which point to missing assumptions or inadequately specified constraints. 
Formally, this yields revised artefacts $(D', R')$ for which satisfaction is re-evaluated, aiming to establish $D'(\theta), S \models R'$ under more precise and defensible assumptions. In this way, automated failure classification informs, but does not replace, human judgement in refining the requirements problem.

\textit{Iteration.}
Because MLS operate under distributional uncertainty and evolving environmental conditions, restoration of requirement satisfaction cannot be assumed to be permanent. A more appropriate notion than static correctness is \emph{resilience}—the ability of a system to maintain, recover, or adapt its behaviour under changing conditions and partial knowledge~\cite{anderssonConceptualFrameworkResilience2020}. REAL therefore adopts an iterative alignment process in which failure identification and multi-layer adaptation are repeatedly applied to the triple $(D,S,R)$. 

Nevertheless, this process does not guarantee convergence in general. New adaptations may expose previously unobserved scenarios, refinements of requirements may alter the effective operational domain, and changes at one layer (e.g.~model-level threshold adjustment) may introduce new violations at another (e.g.~smoothness or efficiency goals). Moreover, because the space of domain instantiations $\Theta$ may be large or effectively unbounded, ensuring universal satisfaction may be infeasible. Iteration thus reflects an ongoing search for alignment rather than a one-time correction.
Understanding the conditions under which this iterative process converges to bounded domain assumptions, monotonic refinements, or probabilistic satisfaction guarantees remains an open research challenge. Formalising convergence criteria and identifying sufficient conditions for ensuring convergence constitute important directions for future work, particularly for safety-critical MLS operating in dynamic environments.

\begin{figure}[t]
\centering
\includegraphics[width=1.0\linewidth]{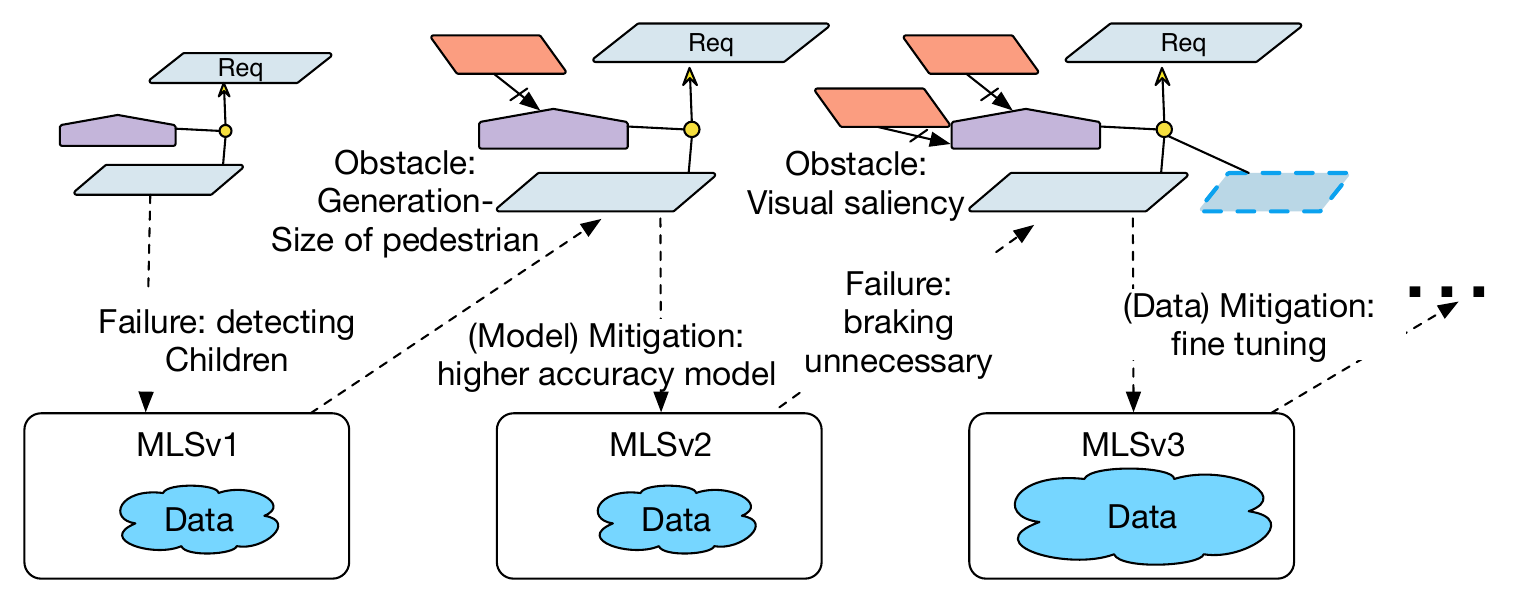}
\caption{Applying REAL to the automated braking system}
\label{fig:mitigate}
\end{figure}

\vspace{0.3em}
\noindent
\textbf{Example.}
In the autonomous braking example, successive MLS versions (MLSv1–MLSv3) illustrate how failures drive obstacle identification and layered mitigation (see Fig.~\ref{fig:mitigate}). The initial implementation (MLSv1) exhibited two dominant failures: unnecessary braking and missed detections of child pedestrians. The latter revealed the obstacle \emph{PedestrianSizeTooSmall}, indicating insufficient representation or sensitivity to small pedestrian instances.
A model-level mitigation (MLSv2) adjusted architecture and decision thresholds, reducing missed detections but increasing unnecessary braking, thereby exposing a trade-off between safety and the soft goal of smooth driving. This showed that model adaptation alone can shift failure modes rather than eliminate them.
A subsequent data-level mitigation (MLSv3) fine-tuned the model with additional child scenarios, improving robustness to size variation but incurring additional data costs. Residual failures linked to clothing and adverse weather highlighted remaining perceptual limitations under challenging environmental conditions.
To address remaining uncertainty, a system-level mitigation introduced confidence-aware cautious braking, reducing safety-critical violations without altering the learned mapping. Finally, requirement-level refinement made explicit detectability constraints and operational assumptions, clarifying the relationship between pedestrian size, visibility, and braking guarantees.
%

\section{Validation}
\label{sec:validation}

This section empirically evaluates REAL on the automated braking example introduced in Section~\ref{sec:example}. 
The objective is to demonstrate how REAL operationalises the satisfaction relation for MLS through (i) grammar-guided scenario exploration, (ii) obstacle-driven failure analysis, and (iii) multi-layer mitigation across data, model, system, and requirements. 

\subsection{Experimental Setup}
\vspace{0.3em}
\noindent\textbf{Simulation Environment (CARLA).}  
We used CARLA~\cite{Carla} to execute autonomous driving scenarios in a high-fidelity simulated urban environment. CARLA enables precise control over ODD parameters including weather, illumination, pedestrian behaviour, and vehicle dynamics. All experiments were executed under reproducible and controlled conditions.

\vspace{0.3em}
\noindent\textbf{Scenario Specification (Scenic).}  
Scenic~\cite{Scenic} was used to define parameterised traffic scenarios enabling systematic variation of structured factors such as pedestrian size and type (adult vs.\ child), crossing direction, distance to vehicle, illumination and weather conditions.

\vspace{0.3em}
\noindent\textbf{Grammatical Evolution Engine.}  
Test scenarios were generated using GRAPE (GRAPE: Grammatical Algorithms in Python for Evolution)~\cite{grape}. 
GRAPE evolves populations of scenario encodings guided by a fitness function that rewards requirement-violating executions (e.g.~collision, late braking, unnecessary braking). We defined a DSL that specifies KAOS constructs using a BNF grammar. We refer the interested reader to the associated \href{https://github.com/darkaengl/REAL}{replication package} for the details of the grammar.


\vspace{0.3em}
\noindent\textbf{Perception Component (YOLOv5).}  
We implemented pedestrian detection using YOLOv5 (variants s and m). 
YOLOv5s provides lower inference latency, while YOLOv5m offers improved detection accuracy at increased computational cost. 
This trade-off is critical for braking performance, where detection latency directly affects stopping distance.

\vspace{0.3em}\noindent\textbf{Datasets.}  
Initial training relied on COCO. Targeted fine-tuning used CARLA-BSP~\cite{wielgosz2023carlabsp} to increase coverage of child pedestrians and adverse visibility conditions.

\vspace{0.3em}\noindent\textbf{Infrastructure.}  
Experiments were executed on a Dell Alienware m18 equipped with Processor
13th Gen Intel® Core™ i9-13980HX (36 MB cache, 24 cores, 32 threads, up to 5.60 GHz Turbo) with fixed software stack.

\subsection{Identifying and Analysing Failure}

This section summarises how we explored requirement satisfaction under structured variations of the operational domain. Each experiment systematically manipulated selected domain parameters while holding others constant, enabling isolation of causal factors contributing to requirement violations. The hardware and software stack remained fixed across all runs to ensure comparability. Table~\ref{tab:experiment_summary} summarises the experimental focus and observed failure modes.

\begin{table}
    \centering
    \caption{Summary of Experiments}
    \label{tab:experiment_summary}
    {\footnotesize
    \begin{tabular}{@{\extracolsep{1pt}}lcc@{}}
       \toprule
Name & Focus  & Observed Failure Mode \\
\midrule
E0 & Baseline & Inconsistent pedestrian detection \\
E1 & Pedestrian size & Systematic failure for children \\
E2 & Distance, direction & Failure at short distance \\
E3 & Visual saliency & Missed detection under brightness variation \\
E4 & Weather & Degraded detection in fog \\
\bottomrule
\end{tabular}}
\end{table}

\vspace{0.3em}\noindent
\textbf{E0: Baseline behaviour.}  
This experiment established baseline performance of YOLOv5s trained on COCO under diverse admissible scenarios.
\textit{Control variables.} Weather was fixed; pedestrian type, clothing, crossing direction, and map location varied across 25 simulation runs.
\textit{Observations.} Detection performance was inconsistent across runs, with unreliable detection of both adults and children. 

\vspace{0.3em}\noindent
\textbf{E1: Pedestrian size (adult vs.\ child).}  
This experiment isolated pedestrian size as a domain parameter to evaluate its impact on safety requirements.
\textit{Control variables.} Weather and clothing contrast were fixed. We explored two pedestrian types: adult and child.
\textit{Observations.} The system consistently failed to detect children with sufficient confidence to trigger braking, whereas adult detection remained comparatively stable. Under admissible domain assumptions, these failures correspond to valid requirement violations rather than out-of-scope scenarios. Obstacle analysis revealed that small pedestrian size constitutes an obstacle to the detection goal.

\vspace{0.3em}\noindent
\textbf{E2: Distance and crossing direction.}  
Given that braking safety depends on reaction time, we evaluated the effect of weather and pedestrian behaviour on detection and stopping behaviour.
\textit{Control variables.} Weather, pedestrian speed, map location, and clothing were fixed. Four variations were explored (long/short distance; left-to-right/right-to-left crossing), with 15 runs per configuration and per pedestrian type.
%
The short-distance adult failure suggested interaction with background clutter, prompting refinement of domain properties related to visual saliency.

\vspace{0.3em}\noindent
\textbf{E3: Clothing variation.}  
This experiment examined whether clothing brightness and contrast affect detection reliability.
\textit{Control variables.} Weather, speed, direction, and map location were fixed. Clothing brightness and contrast were varied for both adult and child models (15 runs each).
\textit{Observations.} Adult detection was generally robust, but at least one bright-clothing case resulted in late braking. Child detection remained unreliable. These results indicate that visibility assumptions embedded in the domain model (e.g.\ \textit{PedestrianVisible}) are sensitive to contrast conditions and require more explicit operationalisation. The failures support grouping under a visual saliency obstacle class.

\vspace{0.3em}\noindent
\textbf{E4: Adverse weather (fog).}  
This experiment assessed detection under degraded visibility.
\textit{Control variables.} Distance, direction, speed, clothing, and map location were fixed. Weather was set to 50\% fog (5 runs per pedestrian type).
\textit{Observations.} Detection confidence decreased substantially and inference latency increased. Children were consistently missed; adults occasionally exhibited delayed detection. These failures highlight a distributional mismatch between training data and fog conditions and reveal that visibility assumptions were insufficiently scoped in the operational design domain. Unlike extreme, unrealistic conditions, 50\% fog remains within plausible deployment contexts, and therefore constitutes a valid requirement-relevant failure mode.

\vspace{0.3em}\noindent
\textbf{Failure grouping and obstacle interpretation.}  
Across experiments, failures clustered into three recurring obstacle patterns: (i) \emph{Size-related detection failures} (children), (ii) \emph{Saliency-related failures} (contrast and clutter), and (iii) \emph{Visibility-related degradation} (fog). These clusters correspond to structured domain properties in the KAOS model and provide actionable abstractions for mitigation. Rather than treating each missed detection as an isolated defect, REAL elevates these empirical counterexamples into obstacle classes that guide targeted adaptation across data, model, system, and requirement layers.

\subsection{Mitigating Failures}

Mitigation operationalises the third phase of the REAL alignment loop (Fig.~\ref{fig:process}) by evaluating alternative repairs across model, data, system, and requirement layers. The objective is to restore requirement satisfaction under admissible domain instantiations while making trade-offs explicit.

\vspace{0.3em}\noindent
\textbf{M1: Model-level adaptation (architecture and thresholds).}  
We compared YOLOv5s and YOLOv5m to assess whether improved detection accuracy restores satisfaction. YOLOv5m increased mean average precision (mAP50-95) from 0.47 to 0.55 (+17\%), but inference latency rose from 5.62\,ms to 7.22\,ms (+28\%). In short-distance scenarios, this latency increase led to delayed braking and persistent violations. 

\vspace{0.3em}\noindent
\textbf{M2: Data-level adaptation (targeted fine-tuning).}  
To address distributional gaps, we fine-tuned YOLOv5s on 1000 annotated child-pedestrian instances derived from CARLA-BSP~\cite{wielgosz2023carlabsp}. Detection confidence for children increased substantially, and collision rate in previously failing scenarios dropped from 100\% to 0\% under nominal weather. Under 50\% fog, successful braking increased from 0\% to 100\%. 

\vspace{0.3em}\noindent
\textbf{M3: Data-level adaptation (few-shot updates).}  
Using 50 manually annotated failure examples (90/5/5 split), limited retraining (8–16 epochs) yielded negligible performance gains ($<5\%$ confidence increase) and did not eliminate child or fog-related failures. Corrective samples proved insufficient to reshape decision boundaries for persistent obstacle classes.

\vspace{0.3em}\noindent
\textbf{M4: System-level adaptation (confidence-aware fail-safe).}  
We introduced proportional braking triggered at 30\% detection confidence, escalating to full braking above 85\%. Child-collision rate decreased from 100\% (baseline) to 8\% (2/25 runs), and adult failures under brightness variation were eliminated (100\% stopping success). 

\vspace{0.3em}\noindent
\textbf{M5: Requirement-level adaptation (explicit scoping and thresholds).}  
Failures revealed underspecified assumptions. We  made explicit: (i) ODD limits for visibility; (ii) quantitative latency and confidence thresholds; and (iii) graded braking guarantees under uncertainty. 

\vspace{0.3em}\noindent
\textbf{Summary.}  
Model scaling improved nominal accuracy (+17\%) but introduced latency-induced violations. Targeted fine-tuning eliminated size- and fog-related failures (up to 100\% improvement). Few-shot learning was ineffective for structural gaps. System-level fail-safe reduced residual collisions by 92\% without retraining. Requirement-level refinement clarified guarantees and bounded satisfaction claims. Overall, cross-layer mitigation reduced child-collision rates from 100\% (baseline) to 8\% and eliminated nominal-weather failures.
These results should not be interpreted as isolating the causal effect of REAL independently from the underlying mitigation techniques. Instead, they show how REAL provides the requirements-level structure for selecting, applying, and evaluating those techniques. Standard ML repair mechanisms, such as fine-tuning, become part of a traceable alignment process: failures are linked to obstacles, obstacles identify the affected domain assumptions or requirements, and mitigation strategies help not only improve the implementation but also its alignment with explicit requirements and domain assumptions.
The results demonstrate that restoring alignment for the automated braking example requires coordinated cross-layer adaptation rather than isolated data augmentation or model optimisation. 

\subsection{Discussion}
\label{sec:discussion}


\noindent
\textbf{Revisiting the research questions.}
The evaluation provides evidence for the three research questions introduced in Section~\ref{sec:intro}. For RQ1, the automated braking experiments show that requirement satisfaction can be operationalised by linking stakeholder goals to executable domain instantiations, simulation traces, admissibility constraints, and observed violations. For RQ2, the study shows that failures discovered through scenario exploration can be systematically reinterpreted as counterexamples and elevated into explicit obstacles, exposing misalignments between requirements, domain assumptions, and system behaviour. For RQ3, the mitigation results show that coordinated adaptation across layers improves requirement satisfaction in the explored scenarios and makes the operational design domain more explicit. These results show that REAL defines a systematic way for practitioners to integrate testing outcomes into RE artefacts. 
The following paragraphs discuss the scope and limits of this evidence.


\vspace{0.3em}
\noindent
\textbf{Lessons learnt and Limitations.}
Our empirical evaluation demonstrates the importance of a systematic RE process in engineering MLS. However, the evaluation remains limited to a simplified safety-critical example with one learned perception component and a restricted set of scenario parameters. In particular, failures involving additional actors or behaviours, such as a pedestrian falling, a dog entering the road, or an interaction between a pedestrian and a cyclist, would require extending the scenario grammar, admissibility constraints, and stakeholder-defined domain assumptions. 
The replication artefact already supports richer scenario structures involving additional actors and behaviours, and REAL treats such extensions as part of the iterative refinement of the requirements and domain assumptions.
Future evaluations should therefore assess how REAL scales when the grammar is expanded to cover multi-actor, interactive, and temporally complex situations.

Although stakeholders are central to REAL, the empirical evaluation did not involve actual stakeholders or practitioners. In this paper, stakeholder input is represented through the initial goals, assumptions, and mitigation choices used in the automated braking example. This was sufficient to demonstrate the mechanics of the framework, but it does not evaluate how stakeholders would negotiate requirements, judge failure acceptability, extend the scenario grammar, or select among competing mitigation strategies in practice. We plan to conduct practitioner studies or participatory evaluations to assess how stakeholders use REAL artefacts to classify failures and refine assumptions and requirements.
In addition, further work is needed to generalise and demonstrate the effectiveness of REAL in other domains, including domains that are not cyber-physical or that involve soft, statistical, fairness, or privacy-related requirements. 

The evaluation shows the importance of using failure as a key concept in requirements modelling and analysis, and highlights how state-of-the-art MLS testing techniques can support this RE process. However, further work is needed to integrate, assess, and contrast different testing methods in REAL. Furthermore, in this paper failure grouping and obstacle identification were performed primarily through manual inspection of traces and experimental conditions. This was sufficient for the controlled example used to explain REAL, but it does not scale to larger scenario spaces and may introduce subjectivity. A natural extension is to automate part of this step by extracting structured features from failure traces and clustering them into recurring failure modes. We refer the interested reader to related work~\cite{dynasto} for a concrete example for clustering failures in autonomous driving. While the resulting failure clusters would not replace stakeholder judgement, they would provide a reproducible and scalable starting point for obstacle analysis.

Finally, our empirical evaluation relied on descriptive mitigation strategies rather than synthesised strategies. There is a wide range of solutions for MLS repair and failure mitigation that we plan to explore and assess in future work.



\vspace{0.3em}\noindent
\textbf{Threats to Validity.}
\textit{Internal validity.} Results depend on simulator fidelity, scenario parametrisation, and stochasticity in scenario generation. CARLA enables controlled experimentation, but may not fully capture real-world perception, weather, and vehicle dynamics. 
\textit{Construct validity.} Failures are measured through braking outcomes and detection confidence; future work should incorporate additional quantitative safety metrics, such as time-to-collision and braking jerk.
\textit{Conclusion validity.} The evaluation provides encouraging evidence that REAL can structure failure discovery, obstacle analysis, and cross-layer mitigation. However, it was not designed to establish superiority over a baseline RE process. Stronger claims require controlled comparisons with repeated runs, fixed test budgets, common safety metrics, and statistical analysis.
\textit{Human judgement.} Failure grouping and obstacle labelling involve manual interpretation, which may introduce subjectivity. Future work should combine stakeholder review with automated trace clustering or pattern mining.
\textit{External validity.} The evaluation focuses on a simplified automated braking example with one learned perception component; generalisation to richer autonomous driving scenarios, multi-component MLS, and non-cyber-physical domains requires further studies.

\vspace{0.3em}\noindent
\textbf{Future work.}
There are three main areas we plan to pursue.
First, while this paper  focused on requirements for MLS, there is increasing emphasis on the role of RE for LLM and foundation models~\cite{Borg24}, particularly from a compliance and regulation perspective~\cite{HassanLRGC00TOL24} or to support software engineering tasks~\cite{Lo24}.
We believe that the idea of focusing on failure described within REAL is worth exploring for LLM and we plan to explore further how REAL can be refined to support RE for LLM.
Second, stakeholders play an important role in REAL, whether for defining the initial goal models, identifying the types of failures, or exploring mitigation strategies. While multiple automation tools are available, some of which based on ML themselves, ultimately, each decision in the RE process involves interaction between humans and autonomous systems~\cite{HuMCSSC22}. 
In fact, collaboration between developers and LLM/MLS is increasingly advocated in software engineering~\cite{Lo23,HassanLRGC00TOL24}. Existing work on human-autonomous systems collaboration~\cite{tosem24} may help support collaboration with stakeholders, especially developers, during the RE process defined in REAL.
%
Third, while this paper focuses on generic failures and stakeholders' needs, values such as fairness are key when engineering MLS~\cite{ShamsBZLW23}. 
We plan to investigate how approaches for modelling and operationalising values, including emotional goal modelling~\cite{HassettBN23} and Values@Runtime~\cite{BennaceurHNZ23}, can be integrated within REAL.
To do so, we plan to revisit the notion of failure itself. Inspired by Edmondson's failure spectrum~\cite{failureBook23}, we will explore how REAL can move beyond a binary view of failure, where a requirement is either satisfied or violated, toward a richer spectrum that accounts for task challenge, uncertainty, and creative exploration. In other words, we aim to redefine failures not only as defects to be repaired, but also as evidence for learning, scoping, and refining requirements, domain assumptions, and mitigation strategies.

\section{Related Work}
\label{sec:relatedWork}

\noindent
\textbf{Requirements Engineering for AI.}
A thorough review of existing work on requirements engineering for MLS is beyond the scope of this paper. In this section, we discuss the approaches most relevant to REAL and refer the interested reader to relevant surveys~\cite{YoshiokaHTCWF21, PeiLWW22, AhmadAABG23, VillamizarEK21}.
Maalej~\textit{et al.}~\cite{MaalejPC23} highlight the need to relate MLS quality to failure types and frequency. They suggest that engineers analyse failures to understand the risks and limitations of MLS and use it to communicate with stakeholders. However, they do not link failure to obstacles and requirements explicitly. 
REAL uses failure to drive the refinement of requirements and domain assumptions.
Andersen and Maalej~\cite{AndersenM24} also proposed different patterns to include human feedback in ML model training and evaluation as well as MLS operation. However, they did not specify any pattern or direction for human feedback for the definition of requirements.
REAL interacts with (human) stakeholders in the analysis of failure and the validation of requirements.

Ahmad~\textit{et al.}~\cite{AhmadAABBG23} propose a conceptual model covering users, models, data, feedback and user control, explainability and trust, and errors and failures. REAL complements this line of work by using failure-driven obstacle analysis and mitigation to refine requirements and domain assumptions.
Anunnaki~\cite{AnuCheng24} relies on goal modelling and obstacle analysis to adapt MLS to the uncertainty of the environment. However, it does not use failure to drive the refinement of the requirements themselves.

\vspace{0.3em}\noindent
\textbf{Industrial guidelines and solutions to Engineering MLS.}
Microsoft Responsible AI Toolbox~\cite{MicrosoftRAI}
includes an error analysis dashboard for identifying model errors and relating it to the relevant data. Engineers using the toolbox can then use other tools for mitigation and updating the models. Similarly, Anthropic Responsible Scaling Policy~\cite{Anthropic} focuses on identifying and mitigating safety failures. 
Google People+AI Guidebook~\cite{GoogleAIGuide} provides a template to engage with stakeholders to define paths forward from failure.
Ahmad\textit{ et al.}~\cite{AhmadAABG23} highlight that while industrial guidelines for responsible and trustworthy AI emphasise identifying errors and linking them to user needs, they still need to be integrated  in the requirements engineering processes for AI. REAL follows these recommendations and proposes an approach to link failure, obstacle analysis, and mitigation strategies.


\vspace{0.3em}\noindent
\textbf{Testing and Safety for MLS.}
The activities of requirements engineering and software testing are closely related, especially when seeking to ensure the quality of software systems, including MLS.
For example, Zhang~\textit{et al.}~\cite{ZhangHML22} highlight the need for requirements analysis when testing MLS.
We briefly discuss some approaches and refer the interested reader to empirical studies on quality issues in MLS~\cite{CoteNBBAK24}. Ashmore~\textit{et al.}~\cite{AshmoreCP21} reviews assurance techniques and their limitations through the ML lifecycle.
However, they assume that the requirements and operating constraints are provided and used to derive the data and model constraints whereas the elicitation and definition of the operating constraints are also dependent on the methods used. In this paper, we consider the intertwining of those operating constraints and the MLS development. 
McDermid~\textit{et al.}~\cite{McDermidCHHJMOPPC24} propose an assurance process that also considers failure. However, it does not use failure to refine requirements or domain assumptions.    
DeepDECS~\cite{CalinescuIMRPSV24} builds upon probabilistic modelling to synthesise controllers that ensure the satisfaction of given requirements for ML components in uncertain environments.
However, it does not consider the uncertainty of the requirements themselves. 
Jordat~\textit{ et al.}~\cite{JodatCNS24} propose to build failure models to explain failures that violate environment assumptions. 
Failures might be due to faults in the MLS or to missing or unknown assumptions.
However, the author do not explain how to use those failure modes to refine requirements. 
By combining Scenic and VerifAI, Viswanadha \textit{et al.}~\cite{ViswanadhaIWKKP21} were able to generate concrete failure scenarios for autonomous driving through testing. 
Those approaches are complementary to ours as they focus on falsification and finding the failure scenarios. The REAL framework does not prescribe a method for finding the failure scenario but rather focuses on using those failures to refine the requirements and domain properties.

\section{Conclusion}
\label{sec:conclusion}
This paper introduced REAL, a failure-driven requirements engineering framework that operationalises requirement satisfaction for MLS. REAL connects stakeholder goals, domain assumptions, and system behaviour through obstacle analysis and multi-layer adaptation. By treating failure as diagnostic evidence rather than a technical defect alone, the framework systematically links scenario-based testing to requirement refinement for MLS.
Our empirical evaluation on an autonomous braking system demonstrates that iteratively identifying, analysing, and mitigating failures makes explicit operational assumptions. The results also show that restoring alignment in MLS requires coordinated cross-layer adaptation rather than isolated model optimisation.
Future work will investigate the generalisability of REAL to other domains and study conditions under which the iterative alignment loop converges. We also plan to extend the framework to better incorporate stakeholder values and a broader notion of failure.

\section*{Data Availability}
A replication package is provided at \href{https://github.com/darkaengl/REAL}{https://github.com/darkaengl/REAL}.
It includes the scenario-generation code, the full BNF grammar, simulation scripts, trained models, and experimental material used in the automated braking example.

\section*{Acknowledgment}
We thank Foutse Khomh and Lionel Briand for the stimulating discussions that helped shape this work. We also thank the reviewers for their valuable feedback. OpenAI ChatGPT was used for language polishing and improving clarity. This work was supported by the Engineering and Physical Sciences Research Council [grant numbers EP/V026747/1, EP/R013144/1]; and Science Foundation Ireland [grant number 13/RC/2094\_P2].

\bibliographystyle{IEEEtran}
\bibliography{refs,se4ml,re4ml2}
\balance

\end{document}